\documentclass[fleqn,twoside]{article}
\usepackage{espcrc2}

\usepackage{graphicx}
\usepackage[figuresright]{rotating}

\def\be{\begin{equation}}
\def\ee{\end{equation}}
\def\bc{\begin{center}}
\def\ec{\end{center}}
\def\bea{\begin{eqnarray}}
\def\eea{\end{eqnarray}}

\def\nn{\nonumber}
\newcommand{\fig}[1]{~{\rm \ref{fig:#1}}}

\newcommand{\AmS}{{\protect\the\textfont2
  A\kern-.1667em\lower.5ex\hbox{M}\kern-.125emS}}

\title{Models of Neutrino Masses and Mixings}

\author{Ferruccio Feruglio\address[MCSD]{Physics Department, University of Padova and I.N.F.N. Padova\\ 
        Via Marzolo 8, Padova, Italy}}
       
\begin{document}

\begin{abstract}
Pocket guide to Terra Incognita of the ``Planisphaerium Neutrinorum'' (http://www.ba.infn.it/\~{}now2004/)
by Eligio Lisi. Includes a survival kit for the $U_{e3}$ territory and 
a first-aid package for the case of
maximal atmospheric mixing angle.
\vspace{1pc}
\end{abstract}

\maketitle

\section{INTRODUCTION}

By looking at the theoretical developements about
neutrino masses and mixing angles of the last few years
we may feel uncomfortable, given the large number of existing 
(and working) models, apparently indicating that, despite the 
formidable experimental progress, no unique and compelling 
theoretical picture has emerged so far.
To some extent this is true, since we cannot speak of a theory 
of neutrino or fermion masses. 
We are still lacking a unifying principle.
In a sense, the current theoretical status of fermion masses
is comparable to that of weak interactions before the 
standard electroweak theory was proposed. 
Several properties of weak processes were well-known, like
universality and the V-A character of the charged current.
Other properties were conjectured, like the existence of
intermediate vector bosons to improve the bad high-energy behaviour
of the theory. But it was only with the introduction of the
unified electroweak theory that renormalizability and the
requirement of SU(2)$\times$ U(1) gauge invariance 
allowed to determine all interactions between fermions and gauge bosons
in terms of only two parameters, the gauge coupling constants of the 
theory.
Unfortunately this is not the case for the interactions between 
fermions 
and spin zero particles, responsible for the fermion spectrum. 
In the electroweak theory, suitably extended to accomodate neutrino oscillations, we may have up
to 22 free observable parameters in the flavour sector! 

In this talk I will summarize some theoretical aspect of 
neutrino masses. I will not attempt to review in detail the many existing 
successful models \cite{af04}, but I will rather address a number of 
questions and open problems such as:
\begin{itemize}
\item[$\bullet$]How many light neutrinos exist?
\item[$\bullet$]Why neutrinos are much lighter than charged fermions?
\item[$\bullet$]Which is the absolute neutrino spectrum favoured by the 
theory?
\item[$\bullet$]What is the most probable range of $U_{e3}$?   
\item[$\bullet$]Is the atmospheric mixing angle $\theta_{23}$ maximal?
\end{itemize}
The tentative answers will illustrate our current ideas
and prejudices, which the future experimental results might
confirm or upset. 

\section{HOW MANY LIGHT NEUTRINOS?}
We know that, besides the result of the LSND experiment, there are 
no independent indications in favour of additional, light, sterile
neutrinos. The existence of a sterile neutrino is disfavoured
by the experiments on both solar and 
atmospheric neutrinos. Even including the uncertainty 
on the standard solar model and in particular on the flux of 
$^8B$-neutrinos, the SNO results imply that the role of sterile
neutrinos in solar oscillations is sub-dominant \cite{bpg}.
Likewise, conversion of atmospheric neutrinos into a sterile
component is disfavoured by several independent properties
of the data, like for instance the zenith
angular dependence of high-energy muons.
When a sterile neutrino is added to the
three active ones, either in the 2+2 or in the 3+1 scheme, the
global fit to oscillation data has a very poor quality. 
More complicated schemes, such as 3+2,
seem to prefer a gap of several eV between the lowest and the highest
levels, which is disfavoured by recent cosmological data from the
cosmic microwave background radiation and large scale structure \cite{cosmo1}.
At the moment three active light neutrinos represent the most
conservative and plausible possibility, but we should of course
wait for an experimental check. If MiniBOONE will confirm the LSND
signal, the working hypothesis of three light neutrinos should be 
abandoned, since not even a CTP violating neutrino spectrum
allows to fit all the data with only three flavours \cite{cpt}.

\section{WHY NEUTRINOS ARE SO LIGHT?}
There are two main theoretical options to describe neutrino masses,
depending on whether the total lepton number L is conserved or not. 
If L is exactly conserved, the smallness of neutrino masses
requires a huge hierarchy of Yukawa couplings within the same
generation ({\em e.g.} $y_e/y_{\nu_e}>10^6$), a fact with no counterpart
in the charged fermion sector, where the corresponding gap is at most 
a factor of about one hundred. Exact conservation of global quantum numbers goes also
against several theoretical prejudices. Baryon and lepton numbers
are broken by anomalies already in the standard model. They are
violated in most of attempts to unify fundamental interactions. Global
quantum numbers are also supposed not to survive the quantization 
of gravity.

\begin{table}[h!]
\caption{Bounds from $0\nu\beta\beta$ searches. Experimental values from ref. \cite{znubb}.}
\label{table1}
\newcommand{\m}{\hphantom{$-$}}
\newcommand{\cc}[1]{\multicolumn{1}{c}{#1}}
\renewcommand{\arraystretch}{1.2} 
\begin{tabular*}{\textheight}{@{}lll}
{\tt experiment} &  $T_{1/2}$ ($\times 10^{25}$ yr)  & $|m_{ee}|$ (eV)\\[2pt]
HM($^{76}Ge$) &  $>1.9$  & $<0.35$ 
\\[2pt]
IGEX($^{76}Ge$)  & $>1.6$  & $<(0.33\div 1.35)$
\\[2pt]
Cuore($^{130}Te$) &  $>5.5\times 10^{-2}$  & $<(0.37\div 1.9)$
\\[2pt]
\end{tabular*}\\[2pt]
\end{table}

If L is conserved, one of the most interesting attempts to describe
neutrino masses is in the context of models with large extra 
dimensions \cite{review}. In such models the right-handed neutrino $\nu_s(x,y)$
is assumed to have access to both the ordinary four space-time dimensions
and to the extra compact dimensions, parametrized by $y$, while all
the other fermions are bound to live in a four-dimensional
slice of the full space-time, parametrized by $x$. All 
four-dimensional modes of the Fourier expansion of
$\nu_s(x,y)$, and in particular the lightest one, $\nu^{(0)}_s(x)$,
come in the low-energy effective theory with the normalization
factor $1/\sqrt{V}$, $V$ being the volume of the extra space: 
\be
\nu_s(x,y)=\frac{1}{\sqrt{V}}\nu^{(0)}_s(x)+...~~~~~.
\label{fourier}
\ee
Therefore the neutrino Yukawa couplings have an extra factor $1/\sqrt{V}$
(in units of a fundamental scale)
compared to the other Yukawa couplings and the smallness of 
the neutrino masses can be explained by the largeness of $V$.
At present, there are no experimental indications supporting
this idea. In principle, some of the higher modes 
$\nu^{(n)}_s(x)$ might take part in neutrino oscillations, 
leading to a peculiar oscillation pattern, but present data 
show that these effects are at most subdominant. 
Moreover, if L is broken by quantum gravity, in these models
the expected effects are huge, since the fundamental scale
of gravity in the higher-dimensional theory is also depleted 
by a factor $1/\sqrt{V}$ compared to the four-dimensional
Planck scale.

\begin{figure}[h!]
$$\hspace{-4mm}
\includegraphics[width=7.0 cm]{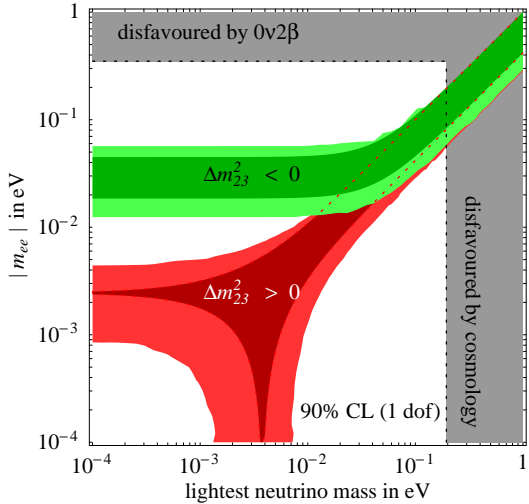}
$$
\caption[]
{\label{fig:postsalt}
From ref. \cite{fsv}, $90\%$ CL range for $|m_{ee}|$
as function of the lightest neutrino mass,
thereby covering all spectra.
The darker regions show how the $|m_{ee}|$ range would shrink if
the present best-fit values of oscillation parameters were confirmed with negligible error.}
\end{figure}

If L is not conserved, then the smallness of neutrino masses
is explained in terms of the largeness of the scale $\Lambda$
at which L is violated:
\be
\frac{(H l_i)(H l_j)}{\Lambda}=\frac{1}{2}\frac{v^2}{\Lambda}\nu_i\nu_j+...
\label{dimfive}
\ee
This scale can be very large, perhaps
as large as the grand unification scale. This makes
the violation of L theoretically very appealing. Neutrino masses
and mixing angles could represent a unique window on otherwise
unaccessible energy scales where the relevant description
might be in terms of a grand unified theory and where
the degrees of freedom required by the see-saw mechanism and by
baryogenesis might become manifest. 

In principle the scale of L violation can also be
much smaller than the electroweak scale, a point of view
opposite to the one adopted above. For instance in supersymmetric
extensions of the standard model with R-parity breaking from
bilinear terms in the superpotential 
\be
W=\mu_i H l_i+... 
\label{Rbreak}
\ee
one neutrino combination becomes massive thanks to a tree-level
contribution proportional to $\mu^2/m_{SUSY}$ ($\mu_i\approx \mu$)
where $m_{SUSY}$
is the supersymmetry breaking scale, expected close to the TeV scale. 
There are two problems with this approach.
First, an ad hoc scale $\mu\approx 0.1$ MeV should be introduced
in our description of particle interactions, with no independent motivation.
Second, the lightest
neutrinos get their masses via radiative corrections so that the typical
hierarchy is too strong to correctly fit $\Delta m^2_{sol}/\Delta m^2_{atm}$
and a further fine-tuning of the parameters is needed.
 
Neutrino oscillations are insensitive to L violation 
and today the most accurate tests involve the search
for neutrinoless double beta decay. The present experiments 
are sensitive to $\vert m_{ee}\vert$, the relevant mass combination,
in the sub-eV range, while future searches will push the 
sensitivity down to about
10 meV, the precise value being unfortunately largely affected
by the theoretical uncertainty coming from nuclear matrix elements.
The effective mass parameter $\vert m_{ee}\vert$ depends 
on neutrino masses, mixing angles and CP-violating Majorana
phases:
\bea
\vert m_{ee}\vert&=&\vert\cos^2\theta_{13}
(\cos^2\theta_{12} m_1+\sin^2\theta_{12} e^{2 i \alpha} m_2)\nn\\
& &+\sin^2\theta_{13} e^{2 i \beta} m_3 \vert
\label{mee}
\eea

Thus the expected range of $\vert m_{ee}\vert$ can be predicted,
as a function of the lightest neutrino mass, from the present knowledge
of the oscillation parameters $(\Delta m^2_{ij},\theta_{ij})$.
The results are shown in fig. \fig{postsalt}. If the neutrino spectrum is
hierarchical with a normal hierarchy, $\vert m_{ee}\vert$ is expected
to be below the sensitivity of the next generation of experiments.
An inverted hierarchy predicts $\vert m_{ee}\vert$ in the 
range ($14\div 53$) meV (90 \% C.L.). If neutrinos are degenerate the present bounds 
on $\vert m_{ee}\vert$ give rise to an upper bound on the common
neutrino mass of about 1 eV, with a large error related to the
poorly known nuclear matrix elements.
Thus the discovery of $0\nu\beta\beta$ decay would not only demonstrate
that lepton number is violated, but it would also largely help in
ordering the neutrino mass spectrum. 

\section{ABSOLUTE SPECTRUM}
The three possible type of neutrino spectrum 
(degenerate, hierarchical with 
normal or inverted hierarchy) are all equally viable at the moment.
For each of them the theory should face specific problems
to achieve a realistic description of the data.

\subsection{Degenerate Spectrum}
There is a converging evidence that such a spectrum is only possible
below the eV scale. The tritium beta decay provides the bound
\be
m_{\nu_e}\equiv
\left(\sum_i |U_{ei}|^2 m_i^2\right)^{1/2} <2.2~ {\rm eV}~(95 \% {\rm C.L.})
\label{tritium}
\ee
where $m_{\nu_e}$ effectively coincides with the common value
$m\approx m_i$ in the degenerate case.
If neutrino masses are of Majorana type, the $0\nu\beta\beta$ bound
$m < 1~ {\rm eV}$ discussed above also applies.
The sum of neutrino masses is strongly limited by recent cosmological
observations and, in the `standard' cosmological model, its   
95 \% {\rm C.L.} upper limit lies in the range between approximately
0.5 eV and 1.5 eV, the exact value depending on the actual set of data
included in the fit \cite{cosmo1,cosmo2}. This remarkable bound depends on
assumptions characterizing the standard cosmological model
and priors in the data analysis, resulting in a large
and difficult to size theoretical uncertainty.
On a more speculative ground, also the elegant idea
of baryogenesis through leptogenesis requires, in its
simplest realization, rather light neutrino masses.
For instance, in the CP-violating, out-of-equilibrium decay
of an heavy right-handed neutrinos (with type I see-saw masses,
hierarchical spectrum of right-handed neutrinos and in the so-called 
strong wash-out regime) the bound $m_i<0.15$ eV holds \cite{leptog1}. By relaxing the
requirement of hierarchical right-handed neutrinos
one can raise the limit up to about 1 eV, but probably not
much more in realistic models \cite{leptog2}. 
Finally, a part of the Heidelberg-Moscow collaboration searching
for $0\nu\beta\beta$ in $^{76}Ge$, claims a positive signal
at the 4$\sigma$ level in their data \cite{klapdor}, that include the full statistics
from 1990 to 2003. Such a signal is compatible with $|m_{ee}|$
in the range $0.1\div 0.9$ eV and favours a degenerate neutrino spectrum.

If neutrinos are quasi degenerate, the type I see-saw mechanism 
\cite{seesaw1,seesaw2}, 
where Dirac and Majorana mass matrices $m_D$ and $M$ are combined
according to the relation:
\be
m_\nu=m_D^T M^{-1} m_D~~~~,
\label{seesaw}
\ee
is of no help. Actually, at first sight it appears quite unplausible 
to get a degenerate spectrum from eq. (\ref{seesaw}),
unless a fine-tuning or a specific symmetry is invoked to relate
$m_D$ and $M$. If we abandon eq. (\ref{seesaw}),
we loose the possible connection between
neutrino masses and the charged fermion masses, which is nicely
present in grand unified theories.

To naturally accommodate a quasi-degenerate neutrino spectrum,
it is convenient to rely on some approximate symmetry,
directly imposed on the five-dimensional operator of eq. 
(\ref{dimfive}). 
For instance, if the lepton 
doublets $l_i$ $(i=1,2,3)$ transform in flavour space as a 
three-dimensional vector of an SO(3) 
rotation symmetry, then in the limit of exact symmetry we have strictly
degenerate neutrinos. The theoretical obstacle to promote this (or a similar)
zeroth-order approximation into a realistic model is that well-established
features of neutrino spectrum, such as the smallness of $\Delta m^2_{sol}/
\Delta m^2_{atm}$, of $\pi/4-\theta_{23}$, of $\pi/4-\theta_{12}$ and of 
$\theta_{13}$
should entirely arise from symmetry breaking effects, since all these
quantities are undetermined in the symmetry limit. At present
there is no general consensus on how to unambiguously 
introduce these breaking effects
in the theory. 

In some model they are simply added by hand and fitted
to the observed parameters, as for instance in the so-called 
`flavour democracy' approach \cite{flavourdem}.
This specific model predicts $\sin^2 2\theta_{23}\approx 0.89$ and is
currently disfavoured by the 2$\sigma$ bound $\sin^2 2\theta_{23}> 0.90$.

In other models the breaking effects are produced by the dynamics of
an underlying symmetry breaking sector. This dynamics is highly non-trivial
and requires a special misalignment between breaking effects in the 
neutrino and in the charged lepton sector, in order to give rise
to the large observed mixing angles \cite{alignment}.

Finally, to avoid a full-fledged theory of breaking terms, we can assume
that the neutrino mass matrix is anarchical (A) \cite{anarchy} with matrix elements
being random, order-one, coefficients in units of a common mass scale $m$:
\be
m_\nu=
\left(
\begin{array}{ccc}
O(1) & O(1) & O(1)\\
O(1) & O(1) & O(1)\\
O(1) & O(1) & O(1)
\end{array}
\right)m~~~~~.
\label{anarchy}
\ee
This speculative scenario has the interesting property that, on a statistical basis, a moderate
hierarchy for $\Delta m^2_{sol}/\Delta m^2_{atm}$ can be produced by the
see-saw mechanism and that large mixing angles are naturally expected.
This is good for the solar and atmospheric angles (although 
$\theta_{23}=\pi/4$ would be accidental), but less good for 
$\theta_{13}$, which is thus expected very close to the present experimental
bound.

\subsection{Inverted Hierarchy}
At the moment the most convincing model for an inverted neutrino 
hierarchy (IH) is the one based on the leading order texture:
\be
m_\nu=
\left(
\begin{array}{ccc}
0 & a & b\\
a & 0 & 0\\
b & 0 & 0
\end{array}
\right)~~~~~,
\label{inverted}
\ee
obtained, for instance, by asking invariance under $L_e-L_\mu-L_\tau$.
Nice features of the above texture are $\theta_{13}=0$ and
$\tan\theta_{23}=-b/a$ (adjustable to 1 by taking $b=-a$). A difficulty arises 
from the prediction of the solar angle, $\theta_{12}=45^0$, 
out by about 6$\sigma$ from the experimental value ($\approx 33^0\pm 2^0$).
\begin{figure}[h!]
\vspace{9pt}
\includegraphics[width=7cm]{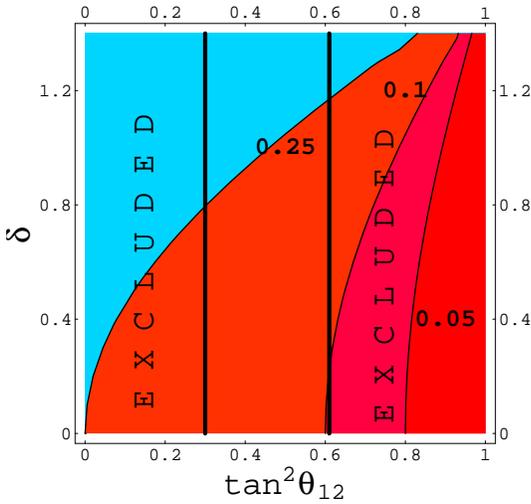}  
\caption{Contours of $|U_{e3}|$ in the plane 
$(\tan^2\theta_{12},\delta_{CP})$. The two vertical lines represent
the present 3$\sigma$ lower and upper limit on $\tan^2\theta_{12}$ 
(see eq. (\ref{tant12})).}
\label{contour}
\end{figure}
We may think to fix this problem by turning on small breaking terms
(filling the zeros in the above texture) also needed to reproduce 
$\Delta m^2_{sol}\ne 0$. However, barring fine-tuning, these breaking terms
obey to the approximate relation $1-\tan^2\theta_{12}\approx 
O(\Delta m^2_{sol}/\Delta m^2_{atm})$, which is off by a factor $>10$ \cite{data}:
\bea
1-\tan^2\theta_{12}&=&(0.39\div 0.70)~(3\sigma)\label{tant12}\\ 
\Delta m^2_{sol}/\Delta m^2_{atm}&=&(0.024\div 0.060)~(3\sigma)~~.
\label{breaking}
\eea  
This difficulty is shared by many models, such as the above mention
flavour democracy model or, more generally, models with a pseudo-Dirac
structure in the 12 sector:
\be
\left(
\begin{array}{cc}
0 & 1\\
1 & 0
\end{array}
\right)~~~~~.
\label{pseudodirac}
\ee
The problem can be solved by taking into account the contribution
to the mixing angles coming from the charged lepton sector \cite{inverse,afm4}.
The mixing matrix is given by $U_{PMNS}=U_e^\dagger U_\nu$,
where $U_\nu$ and $U_e$ come from the diagonalization of the 
neutrino and charged lepton mass matrix, respectively. 
By adopting a standard parametrization for 
$U_{\nu,e}=U^{\nu,e}_{23}\cdot U^{\nu,e}_{13}\cdot
U^{\nu,e}_{12}$, we can absorb $U^e_{23}$ into $U_\nu$ 
and, by assuming $\theta^\nu_{23}=\pi/4$, we have a 
bimaximal mixing in the neutrino sector, to start with.
Then we can compute the effect of small angles
$\theta^e_{12}$ and $\theta^e_{13}$ on $U_{PMNS}$, obtaining:
\bea
1-\tan^2\theta_{12}&=&2 \sqrt{2}~ Re(u+v)\\
|U_{e3}|&=&\frac{1}{\sqrt{2}} |u-v|\\
\delta_{CP}&=&arg(u-v)\\
\tan^2\theta_{23}&=&1
\label{charged}
\eea
where $u$ and $v$ are complex numbers,
$(|u|,|v|)\equiv (\sin\theta^e_{12},\sin\theta^e_{13})$
and the above results hold to first order in $(|u|,|v|)\ll 1$.
By analogy with the quark sector, we may tentatively assume:
\be
|v|\ll |u|\approx \lambda\approx 0.22~~~~~,
\ee
and we get:
\bea
|U_{e3}|&\approx&\frac{(1-\tan^2\theta_{12})}{4\cos\delta_{CP}}\label{corr}\\
1-\tan^2\theta_{12}&\approx& 2 \sqrt{2}~\lambda\approx0.6~~~~~.
\eea
We see that $\theta_{12}$ is indeed corrected by the right amount (also suggesting
an intriguing complementarity between $\theta_{12}$ and $\lambda$ \cite{complementarity}) but,
at the same time, a non-vanishing $|U_{e3}|$ is produced. Actually,
from equation (\ref{corr}), a rather
large value of $|U_{e3}|$ is expected. 
In fig. \ref{contour} a contour plot
of $|U_{e3}|$ in the plane $(\tan^2\theta_{12},\delta_{CP})$ shows
that in most of the available parameter space, where we keep assuming
$|v|\ll |u|$, $|U_{e3}|$ is larger that 0.1. 
This model of inverse hierarchy can be realized both within and without
the see-saw mechanism. 

\subsection{Normal Hierarchy}
The interesting feature of a hierarchical neutrino spectrum with
normal hierarchy is the central role played by the see-saw mechanism,
at variance with inverse hierarchy where such a mechanism is an option
and with the degenerate case where it may even represent an obstacle.
In the so-called flavour basis, where the charged leptons are diagonal,
the condition for a hierarchical neutrino spectrum and large atmospheric
and solar mixing angles is that the 23 block of the neutrino mass matrix
has comparable matrix elements with a suppressed
determinant. In the context of the see-saw mechanism there are two
natural possibilities to met this condition.
\begin{itemize}
\item[$\bullet$]
The see-saw mechanism is dominated by a single, relatively `light'
right-handed neutrino, equally coupled to $\nu_\mu$ and $\nu_\tau$ \cite{dominance};
\item[$\bullet$]
The mass matrices $m_e$ and/or $m_D$ (with the convention 
($\bar{R} m_{e/D} L$) have an approximate `lopsided' structure \cite{lopsided}:
\be
\left(
\begin{array}{ccc}
0 & 0 & 0\\
0 & 0 & 0\\
0 & a & b
\end{array}
\right)~~~~~.
\label{lops}
\ee
\end{itemize}
A typical texture for neutrinos with normal hierarchy, as
arising in models with U(1) flavour symmetry, is
\be
m_\nu=
\left(
\begin{array}{ccc}
\epsilon^2 & \epsilon & \epsilon\\
\epsilon & 1 & 1\\
\epsilon & 1 & 1
\end{array}
\right)m~~~~~,
\label{nortext}
\ee
where $\epsilon<1$ is the symmetry breaking parameter and all entries are 
given up to unknown O(1) coefficients.
The angle $\theta_{13}$ is of order $\epsilon$. A hierarchical spectrum and 
a large solar mixing angle require that the determinant of the 23 block is 
of order $\epsilon$. 
This may happen accidentally, and we may call the corresponding 
models semi-anarchical (SA),
or thanks to one of the two possibilities mentioned above (H models).
Both these possibilities can be realized in the context of a grand unified 
theory, where they are compatible with the observed smallness of the quark
mixing. In particular, the large mixing angle among charged fermions
expected in the lopsided texture corresponds in SU(5) to a mixing between 
right-handed down quarks and, as such, is not directly manifest in weak 
processes. 

\section{ESTIMATES OF $U_{e3}$}
It is quite difficult to arrange for a tiny $U_{e3}$ in a realistic
model of neutrino masses. By tiny here I mean well below 0.05,
which approximately corresponds to the square $\lambda^2$ of the Cabibbo angle. 
The range between 0.05 and the present upper bound
will be presumably explored in about ten years from now and it will
not require a neutrino factory program.
\begin{figure}[h!]
\vspace{9pt}
$$
\includegraphics[width=7cm]{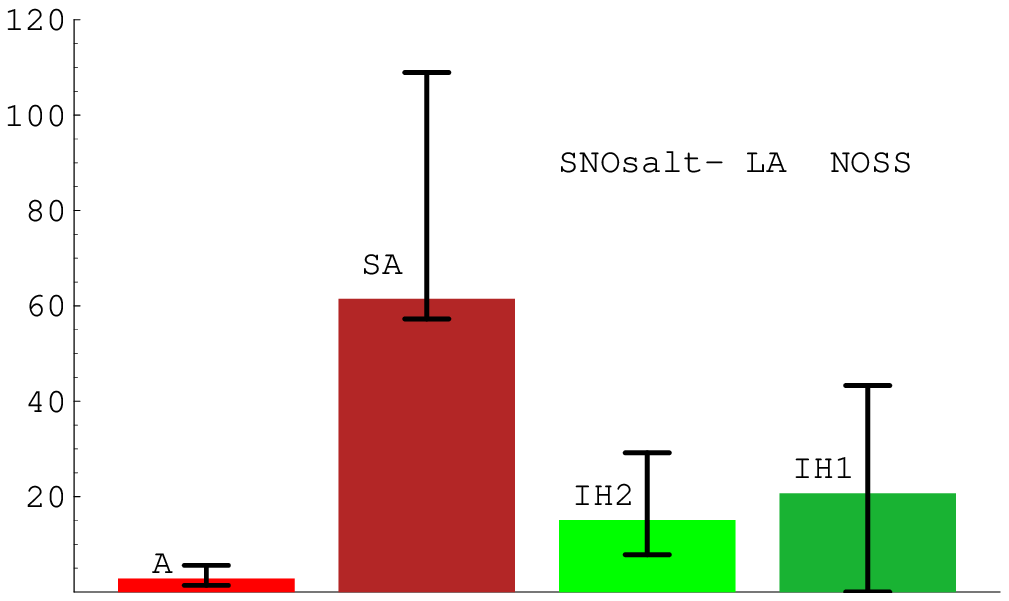}  
$$
\includegraphics[width=7cm]{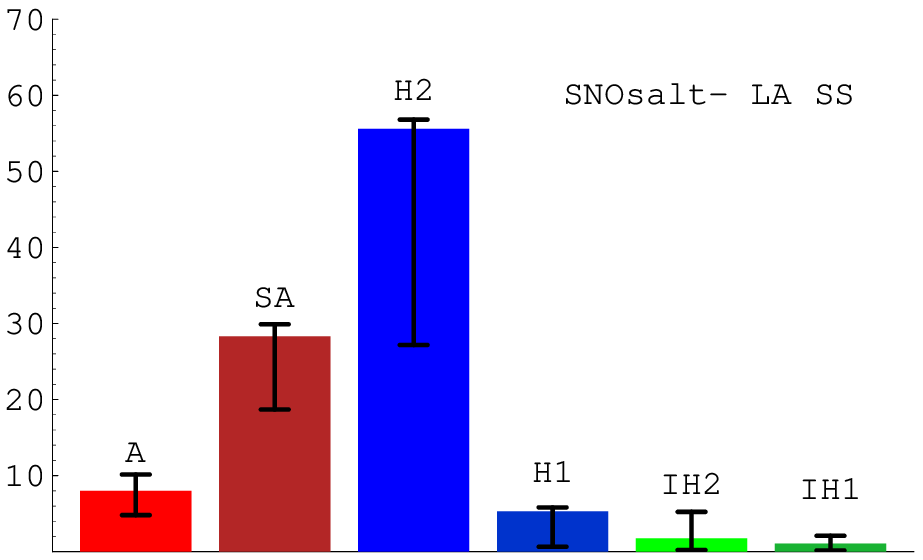}
\caption{Relative success rates of models with U(1) flavour symmetry,
analyzed both without (NOSS) and within (SS) the see-saw mechanism. 
The sum of the rates has been normalized to 100. The results correspond to the 
optimized values of $\epsilon$: $1$, $0.2$, $0.25$, $0.25$, 
for the models A, SA, IH2, IH1 (NOSS case) and $1$, $0.25$, $0.12$, $0.2$, $0.45$, $0.25$ for
the models A, SA, H2, H1, IH2, IH1 (SS case). 
The error bars represent the linear sum of the systematic error 
and the statistical error. For details, see ref. \cite{afm3}.}
\label{rates}
\end{figure}

We have already seen that in inverted hierarchy the leading order
prediction $U_{e3}=0$ is substantially modified by the corrections
coming from the charged lepton sector, so that, barring cancellation,
we expect (see eqs. (\ref{tant12},\ref{corr})):
\be
|U_{e3}|>\frac{(1-\tan^2\theta_{12})}{4}\approx(0.10\div 0.18)~[3\sigma]
~~,
\ee
which is within a factor of two from the present bound.
For a normal hierarchy, we may estimate $U_{e3}$ from the 
approximate relation
\be
U_{e3}\approx \sin\theta^\nu_{13}-\sin\theta_{23}\sin\theta^e_{12}~~~~~,
\label{ue3}
\ee
which holds under the plausible assumptions that both the 
contributions from the
neutrino and the charged lepton sectors are small ({\em i.e.} no 
cancellations between large terms) and that $\sin\theta^e_{12}$
dominates over $\sin\theta^e_{13}$.
In `typical' models $\sin\theta^\nu_{13}$ is of order 
$\sin\theta_{12}\sqrt{\Delta m^2_{sol}/\Delta m^2_{atm}}$,
which lies in the range $(0.03\div 0.3)$
(by allowing for a 3$\sigma$ experimental uncertainty
and by modeling through factors 1/2 and 2 the theoretical uncertainty,
here and in the following estimates). 
The contribution from the charged leptons is typically
of order $\sin\theta_{23}\sqrt{m_e/m_\mu}\approx (0.02\div 0.1)$.
Again, excluding cancellations between the two contributions 
we find that $U_{e3}$ cannot be tiny.
(If we used the analogue of eq. 
(\ref{ue3}) to estimate $V_{ub}$ from $V_{cb}\approx \lambda^2$ and 
$V_{us}\approx \lambda$, we would get $V_{ub}\approx \lambda^3$.)
In models of degenerate neutrinos everything is determined by
breaking terms and we should proceed through examples. 
For instance, in a model where $\Delta m^2_{sol}$ and $U_{e3}$
are determined by renormalization group evolution, 
$U_{e3}\approx \sqrt{\Delta m^2_{sol}/2 \Delta m^2_{atm}}\approx 
(0.04\div 0.4)$. In the flavour democracy model, 
$U_{e3}\approx\sqrt{2 m_e/3 m_\mu}\approx (0.03\div 0.1)$.
In the anarchical approach, $U_{e3}\approx O(1)$.

In a specific class of models a more quantitative analysis
can be performed. For instance, models with a U(1) flavour
symmetry are flexible enough to cover all possible
type of neutrino spectrum. By varying the U(1) charge assignment
we can get neutrino mass matrices of the following structure:
\bea
m_\nu&=&
\left(
\begin{array}{ccc}
\epsilon^2 & \epsilon & \epsilon\\
\epsilon & 1 & 1\\
\epsilon & 1 & 1
\end{array}
\right)m~~~~~
\label{normal}
\\
&{\rm or}& \nn\\
m_\nu&=&
\left(
\begin{array}{ccc}
\epsilon^2 & 1 & 1\\
1 & \epsilon^2 & \epsilon^2\\
1 & \epsilon^2 & \epsilon^2
\end{array}
\right)m~~~~~,
\label{invert}
\eea
where all matrix elements are up to unknown O(1) coefficients,
which the U(1) symmetry is unable to predict, and $\epsilon$ is a free
parameter. If $\epsilon=1$, these textures reproduce the A
model. Otherwise, $\epsilon<1$ represents the order parameter of U(1)
symmetry breaking. In this case, eq. (\ref{normal}) gives rise to both
the SA and the H models, depending on whether the determinant of the
23 block is typically of order 1 or $\epsilon$, respectively. Eq.
(\ref{invert}) corresponds to the IH model.
Given the presence of unknown O(1) parameters, these textures can only
predict masses and mixing angles on a statistical basis \cite{afm3}.
In practice the O(1) parameters are taken as complex random numbers,
with a given initial distribution. The corresponding predictions are
compared with the 3$\sigma$ experimental windows:
\bea
0.018<&\vert\Delta m^2_{sol}/\Delta m^2_{atm}\vert&<0.053\nn\\
0.30<&\tan^2\theta_{12}&<0.64\nn\\
0.45<&\tan^2\theta_{23}&<2.57\nn\\
&|U_{e3}|&<0.23
\label{windows}~~~~~,
\eea
and a success rate is evaluated. The procedure is iterated
and the parameter $\epsilon$ is optimized to maximize the success
rate of each model. In fig. \ref{rates} the results of such an optimization are
displayed and we can see that models possessing some degree
of order are preferred over models based on structure-less mass matrices.
Also, for SA, H, and IH models, an expansion
parameter $\epsilon$ numerically close to the Cabibbo angle
is needed (see figure caption) to better account for the smallness of 
$\Delta m^2_{sol}/\Delta m^2_{atm}$ and of $|U_{e3}|$.
\begin{figure}[h!]
\vspace{9pt}
$$
\includegraphics[width=7cm]{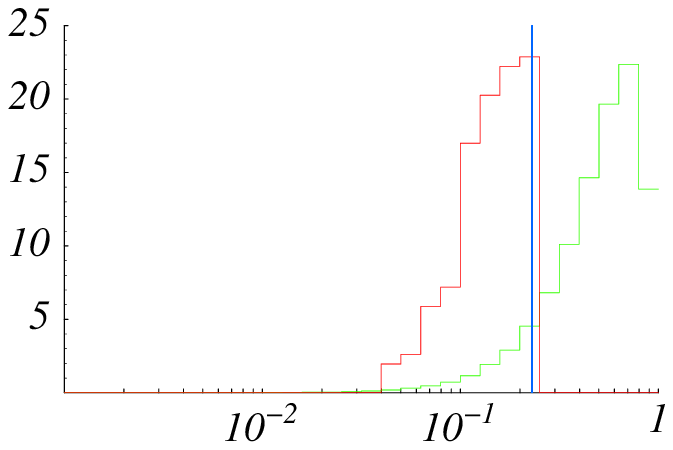}  
$$
\includegraphics[width=7cm]{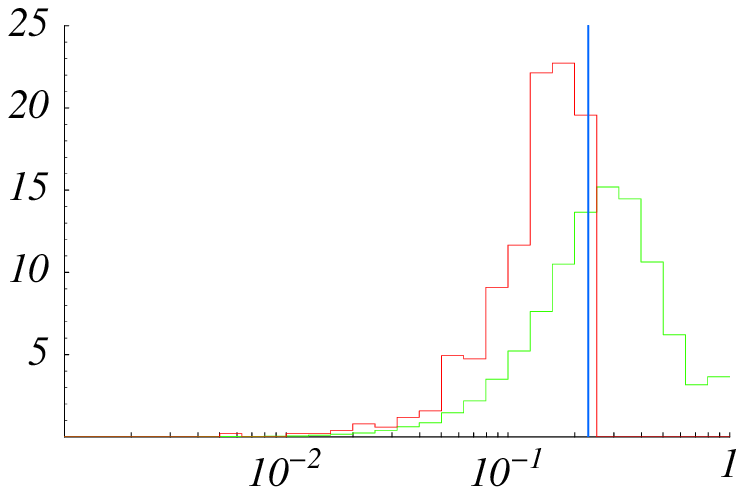}
\caption{Distribution of $|U_{e3}|$ \cite{isa} for the models A (top) and SA (bottom), with see-saw mechanism,
before (green, light) and after (red, dark) the selection made by the experimental cuts in eq. 
(\ref{windows}).
The vertical line shows the present experimental bound.}
\label{ue3asa}
\end{figure}
\begin{figure}[h!]
\vspace{9pt}
$$
\includegraphics[width=7cm]{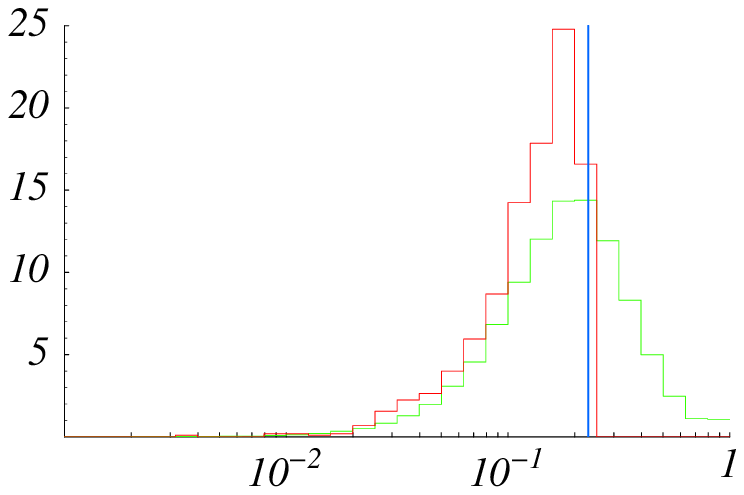}  
$$
\includegraphics[width=7cm]{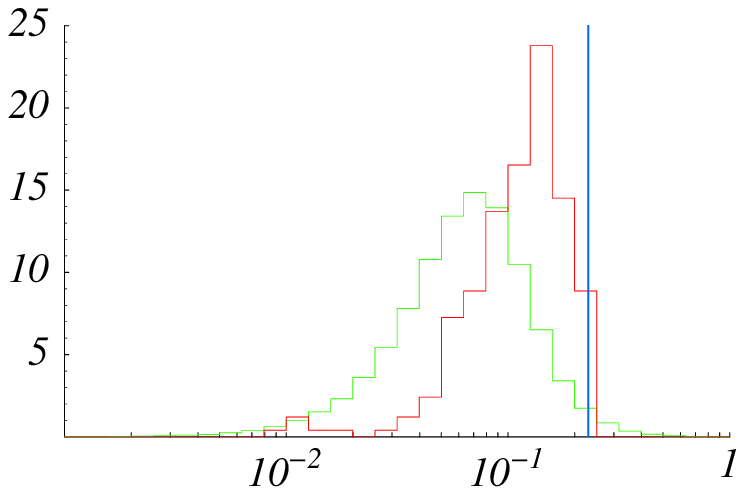}
\caption{Distribution of $|U_{e3}|$ \cite{isa} for the models H2 with see-saw mechanism (top) and IH without see-saw 
mechanism (bottom), 
before (green, light) and after (red, dark) the selection made by the experimental cuts in eq. 
(\ref{windows}). The vertical line shows the present experimental bound.}
\label{ue3h2ih}
\end{figure}
For the successful models we can finally
analyze the distributions of $|U_{e3}|$ \cite{isa}, some of which are displayed in
fig. \ref{ue3asa} and fig. \ref{ue3h2ih}. Such distributions, which clearly have just the purpose
of providing a rough orientation, show that a value of $|U_{e3}|$
close to the present experimental upper bound is preferred, with
`statistical' tails extending down to, but not below, the percent
level. These tails are shorter for the A and IH models.

If `normal' models are likely to predict $|U_{e3}|$ at or above the
$\lambda^2$ level, we may ask if `special' models exist, where
$|U_{e3}|\ll \lambda^2$. First we can analyze the case of models
where a small $|U_{e3}|$ does not arise from the cancellation of 
comparable contributions. In this case $|U_{e3}|$ is approximately given by
eq. (\ref{ue3}) and both $\theta^\nu_{13}\ll \lambda^2$ and 
$\theta^e_{12}\ll \lambda^2$ are required in order to obtain a
tiny $|U_{e3}|$. It is relatively easy to get $\theta^\nu_{13}=0$
in a symmetry limit. For instance, the textures in eq. 
(\ref{normal},\ref{invert}), implied by U(1) flavour symmetries,
predict $\theta^\nu_{13}=0$ when the symmetry 
breaking parameter
$\epsilon$ is set to zero. However, turning $\epsilon$ on,
we get $\theta^\nu_{13}\approx 
\sqrt{\Delta m^2_{sol}/\Delta m^2_{atm}}$ and 
$\theta^\nu_{13}\approx 
\Delta m^2_{sol}/\Delta m^2_{atm}$, for (\ref{normal}) and (\ref{invert}),
respectively. For a more efficient suppression of $\theta^\nu_{13}$
we can look for a symmetry that guarantees $\theta^\nu_{13}=0$ and that, 
in the neutrino sector, does not need any correction term.
For instance a $Z_2$ symmetry exchanging the muon and tau SU(2) doublets 
implies:
\be
m_\nu=
\left(
\begin{array}{ccc}
x & y & y\\
y & w & z\\
y & z & w
\end{array}
\right)~~~~~,
\label{grimus}
\ee
which in turn leads to $\theta^\nu_{23}=\pi/4$ and $\theta^\nu_{13}=0$
and allows to fit $\theta^\nu_{12}$ and $\Delta m^2_{sol}/\Delta m^2_{atm}$
by adjusting the free parameters $x$, $y$, $z$ and $w$.
However, in realistic models such a symmetry is expected to be broken 
by O(1) effects in the charged lepton sector 
to account for $m_\mu\ll m_\tau$. Moreover, in general it is  
difficult to accommodate $\theta^e_{12}\ll \lambda^2$ at the same
time. As a consequence the previous estimates of $U_{e3}$ probably 
apply also to this class of models \cite{moha}. 

Alternatively, $|U_{e3}|\ll \lambda^2$ can occur as the result of
a compensation between O(1) terms. This may happen in models where
the lepton mixing angles come entirely from the charged lepton sector
and $U_{e3}$ is given by:
\be
U_{e3}=\sin\theta^e_{12}\sin\theta^e_{23}-\cos\theta^e_{12}
\sin\theta^e_{13}\cos\theta^e_{23}~~~~~.
\ee
In a particular model of this class it is indeed possible to arrange
$U_{e3}\approx m_e/m_\tau\approx \lambda^4$, even though the 
construction is considerably more complicated than in `normal'
models \cite{afm4}. 

In summary, it is reasonable to expect that most of the plausible
range for $|U_{e3}|$ will be explored in about ten years from now.
Without further information, discovery of $|U_{e3}|$ in the range
$(0.10\div 0.23)$ will not favour a particular model and/or a 
particular type of spectrum. A relatively small, but still sizeable
$|U_{e3}|$, between 0.05 and 0.10, is difficult to reconcile with anarchy and 
with inverse hierarchy, in its simplest version. 
If $|U_{e3}|\ll 0.05$ then a very narrow,
but not empty, subset of the existing models would be selected. 
The interested reader can compare the above estimates with those
of the recent literature \cite{ue3}.

\section{IS $\theta_{23}$ MAXIMAL ?}
The present best value of $\theta_{23}$ is very close to $45^0$,
with a still large 3$\sigma$ uncertainty:
\be
35.7^0<\theta_{23}<55.6^0~~~.
\label{range23}
\ee
Since the determination of $\theta_{23}$ is mainly due to the measurement of the $\nu_\mu$
survival probability and $\delta P_{\mu\mu}\approx \delta(\sin^2 2\theta_{23})\approx
4\delta(\Delta^2)$ $(\Delta\equiv \theta_{23}-\pi/4)$,
it will be very difficult in the future to improve the experimental precision on $\theta_{23}$.
To achieve a reduction of the error by a factor of two, we will probably need superbeams and
a time scale of about ten years \cite{lindner}.
Nevertheless future data might confirm the preference for a maximal $\theta_{23}$
and motivate the search for a natural explanation of such a peculiar feature
of lepton mixing. 
The natural question that comes to mind is if $\theta_{23}=\pi/4$ can be understood in some symmetry
limit of the flavour sector. If true, this could be of great help in our search for a unifying
principle. Several observable quantities can be understood via exact or approximate symmetries
of particle interactions, as for example the photon mass whose vanishing is guaranteed by
the U(1) electromagnetic gauge invariance or as the electroweak $\rho$ parameter, 
which an approximate custodial symmetry requires to be close to one.

I will conclude this presentation by illustrating a perhaps not well-known and to some extent counterintuitive property.
Contrary to expectations, a maximal mixing angle $\theta_{23}$ can never arise in a symmetry limit
of whatever flavour symmetry (global or local, continuous or discrete), provided such
a symmetry is only broken by small effects in the real world, as we expect for a meaningful
symmetry. This result does not necessarily have a negative impact on the theory.
The immediate implication is that, in the context of a sensible flavour symmetry,
$\theta_{23}=\pi/4$ can only arise as a breaking effect. Therefore, if really $\theta_{23}$
is maximal or nearly maximal in nature, our description of lepton mixing will be severely constrained
and in any model of maximal atmospheric mixing based on a symmetry approach,
both the underlying symmetry and the symmetry breaking sector should be carefully
chosen. 

I give here a shortcut to the full argument. Consider the charged lepton mass matrix:
\be
m_l=m^0_l+...~~~,
\label{chargedleptons}
\ee
where $m^0_l$ represents the limit of an exact flavour symmetry F and dots are the symmetry breaking
effects that vanish when F is exact. We should distinguish two cases.
The first case is when $m^0_l$ has rank less or equal to one. This is true for most
of realistic flavour symmetries. In a suitable basis:
\be
m^0_l=
\left(
\begin{array}{ccc}
0 & 0 & 0\\
0 & 0 & 0\\
0 & 0 & m_\tau
\end{array}
\right)~~~~~,
\label{rankone}
\ee
and only the $\tau$ lepton possibly has a non-vanishing mass in the symmetry limit.
Electron and muon masses arise as symmetry breaking effects.
It is clear from eq. (\ref{rankone}) that $\theta^e_{12}$ is undetermined in this limit.
Therefore 
\be
\tan\theta_{23}=\tan\theta^\nu_{23}\cos\theta^e_{12}+\frac{\tan\theta^\nu_{13}}{\cos\theta^\nu_{23}}
\sin\theta^e_{12}~~~~~
\label{tt23}
\ee
is also undetermined, no matter what is the leading neutrino mass matrix. When small breaking terms 
are turned on, both $m_l^0$ and $m_\nu^0$ are modified and, to obtain a maximal atmospheric mixing angle,
we need a conspiracy between the breaking effects in the neutrino and in the charged lepton sector.
For instance, in the limit of an exact SO(3) flavour symmetry under which the lepton doublets transform
as a vector, we can have:
\be
m^0_\nu=
\left(
\begin{array}{ccc}
1 & 0 & 0\\
0 & 1 & 0\\
0 & 0 & 1
\end{array}
\right)m ~~~~~,
\label{so30nu}
\ee
\be
m^0_l=
\left(
\begin{array}{ccc}
0 & 0 & 0\\
0 & 0 & 0\\
0 & 0 & 0
\end{array}
\right) ~~~~~,
\label{so30l}
\ee
and a possible breaking giving rise to a maximal $\theta_{23}$ is described by
small terms $x,x',y,y'\ll 1$ such that:
\be
m^0_\nu=
\left(
\begin{array}{ccc}
1 & 0 & 0\\
0 & 1+x' & 0\\
0 & 0 & 1+x
\end{array}
\right)m ~~~~~,
\label{so3nu}
\ee
\be
m^0_l=
\left(
\begin{array}{ccc}
0 & 0 & 0\\
0 & y' & -y'\\
0 & y & y
\end{array}
\right)\frac{v}{\sqrt{2}} ~~~~~.
\label{so3l}
\ee
The second possibility is that $m^0_l$ has rank larger or equal than two:
\be
m^0_l=
\left(
\begin{array}{ccc}
0 & 0 & 0\\
0 & m_\mu & 0\\
0 & 0 & m_\tau
\end{array}
\right)~~~~~.
\label{ranktwo}
\ee
In this case $\theta_{23}=\pi/4$ is possible in the symmetry limit, but we are left with the problem
of explaining why $m_\mu\ll m_\tau$. Since both masses are non-vanishing in the symmetry
limit, generically large O(1) breaking effects are needed to correct the charged lepton spectrum.
This is precisely the kind of symmetry we have excluded from our considerations since
O(1) corrections can easily upset the result $\theta_{23}=\pi/4$.

\section{CONCLUSION}
Many questions about the nature of the neutrino mass spectrum are still unanswered.
The most conservative theoretical approach is in terms of three light neutrinos
whose lightness is naturally described by L violation at a 
large energy scale. The absolute spectrum is an open experimental problem,
though several theoretical (and speculative) properties like type I see-saw, leptogenesis, relationship with
other fermion masses in GUTs favour, in their simplest realization, a hierarchical spectrum.
At present both normal and inverse hierarchy are equally viable, but in the future
tests of $\tan^2\theta_{12}$ and $|U_{e3}|$ might constrain more strongly
the inverse hierarchy scheme. In most of known models $|U_{e3}|$ is not much smaller
than the square of the Cabibbo angle. A tiny $|U_{e3}|$ is possible in principle
but needs the construction of an ad hoc model. It is true that the hierarchies among observed
quantities are less pronounced in the lepton than in the quark sector. Nevertheless
an expansion parameter numerically close to the Cabibbo angle largely
helps in reproducing the relative smallness of $\Delta m^2_{sol}/\Delta m^2_{atm}$
and of $|U_{e3}|$. Finally a nearly maximal $\theta_{23}$ represents a perhaps premature,
but quite interesting challenge, which would strongly constrain the present theoretical
framework.

\section*{Acknowledgment}
I am deeply indebted to Guido Altarelli, for 
transmitting his passion for neutrinos to me.
I am also grateful to Isabella Masina for her
valuable help in preparing this talk. I thank both of them for the
very pleasent collaboration on the topics of this talk.


\begin{thebibliography}{9}
\bibitem{isa}
These distributions have been kindly worked out 
by Isabella Masina.
\end{thebibliography}
\end{document}